\documentstyle[pra,aps,eqsecnum,epsfig,multicol,amsmath]{revtex}

\def\figone{\begin{figure}
  \begin{center}
   \mbox{\epsfig{file=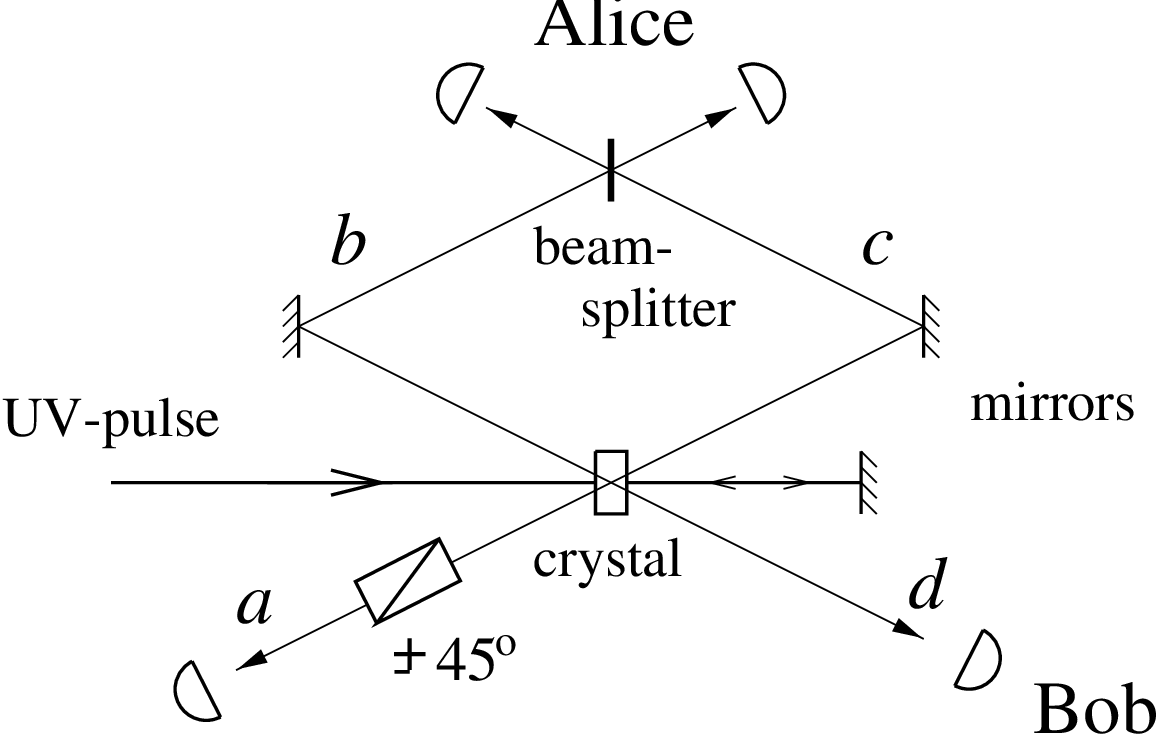, height=3.7cm, angle=270}}
  \end{center}
  \end{figure}
  {\small Fig.\ 1. Schematic representation of the experiment conducted in 
	Innsbruck. A {\sc uv}-pulse is sent into a non-linear crystal, thus 
	creating an entangled photon-pair. The {\sc uv}-pulse is reflected by a
	mirror and returned into the crystal again. This reflected pulse 
	creates the second photon-pair. Photons $b$ and $c$ are sent into a 
	beam-splitter and are detected. This is the Bell measurement. Photon 
	$a$ is detected to prepare the input state and photon $d$ is the 
	teleported output state Bob receives. In order to rule out the 
	possibility that there are no photons in mode $d$, Bob detects this 
	mode.}\medskip
 \label{fig:innsbruck}
}

\def\figtwo{\begin{figure}
  \begin{center}
   \mbox{\epsfig{file=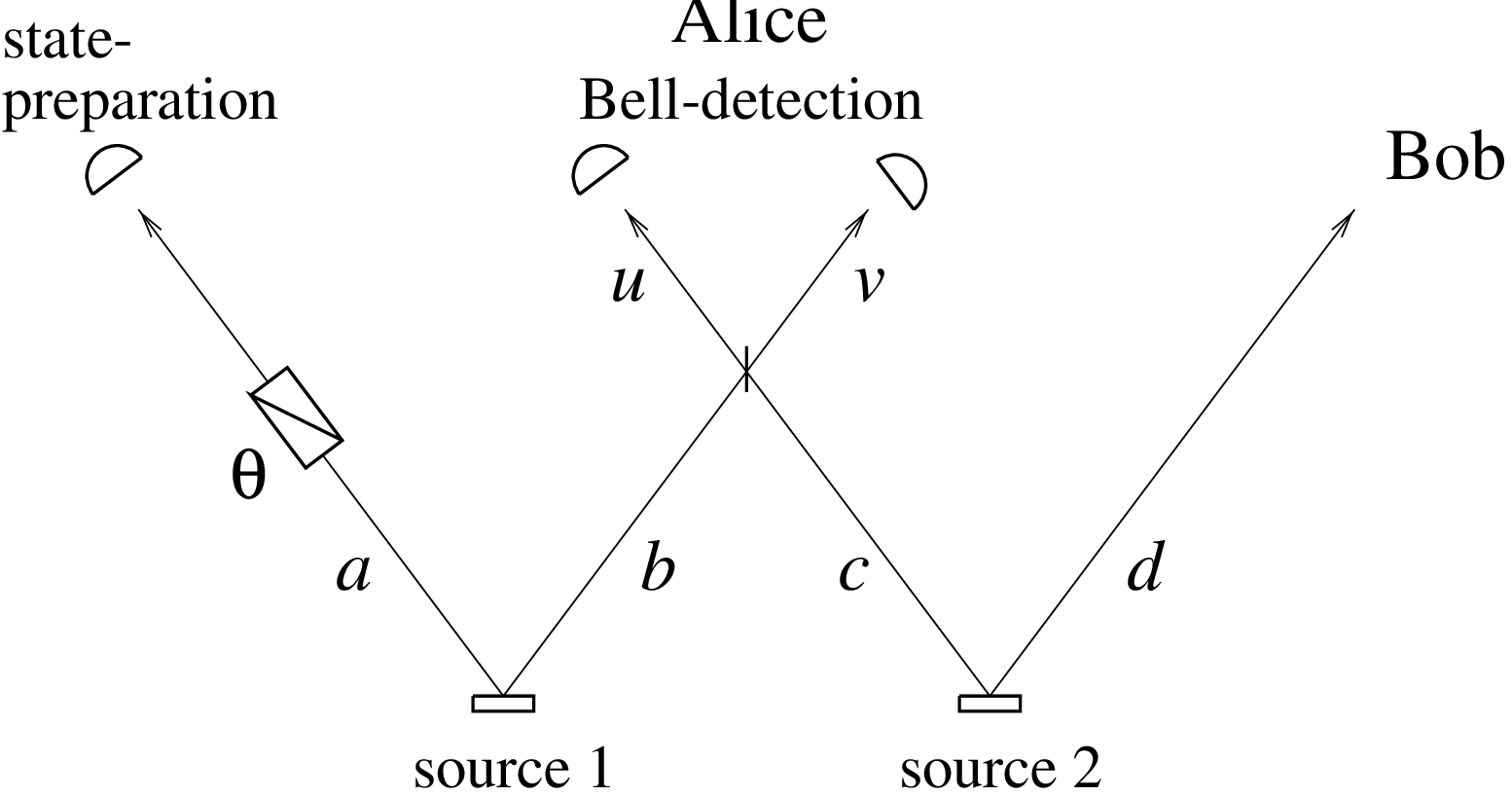, height=3cm, angle=270}}
  \end{center}
  \end{figure}
  {\small Fig.\ 2. Schematic `unfolded' representation of the teleportation 
	experiment with two independent down-converters and a polarisation 
	rotation in mode $a$. The state-preparation detector is actually a 
	detector cascade and Bob does not detect the mode he receives.}\medskip
  \label{model}
}

\def\figthree{\begin{figure}
  \begin{center}
   \mbox{\epsfig{file=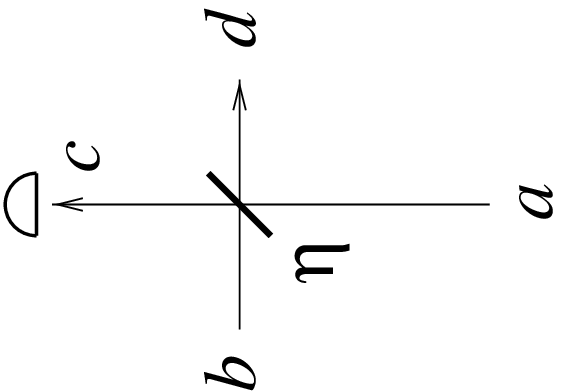,height=2.5cm,angle=270}}
  \end{center}
  \end{figure}
  {\small Fig.\ 3. A model of an inefficient detector. The beam-splitter will 
	reflect part of the incoming mode $a$ to mode $d$, which is thrown 
	away. The transmitted part $c$ will be sent into a ideal detector. 
	Mode $b$ is vacuum.}\medskip
  \label{fig:detector}
}

\def\figfour{\begin{figure}
  \begin{center}
   \mbox{\epsfig{file=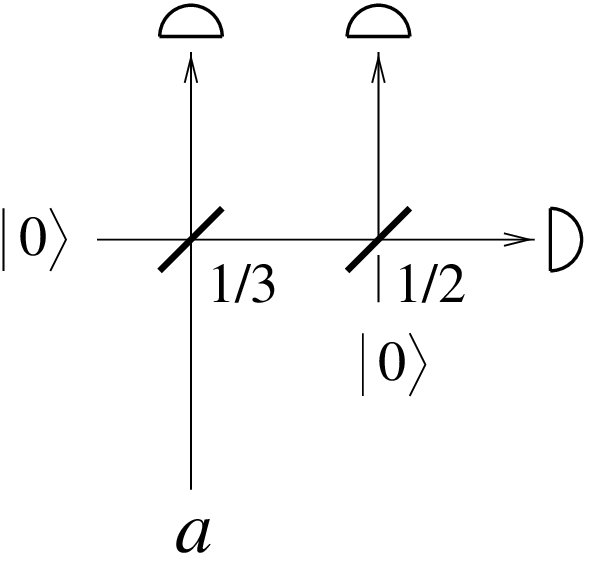,height=2.5cm,angle=270}}
  \end{center}
  \end{figure}
  {\small Fig.\ 4. A simple detector cascade. The fractions $1/2$ and $1/3$ 
	are the beam-splitter's intensity transmission coefficients. Several 
	photons in mode $a$ are likely to enter different detectors, thus 
	revealing that more than one photon was present in this mode.}\medskip
  \label{fig:cascade}
}

\begin{document}

\title{Post-Selected versus Non-Post-Selected Quantum Teleportation \\ using 
	Parametric Down-Conversion}
\author{Pieter Kok\cite{pieter}$^{1}$ and Samuel L.\ Braunstein$^{1,2}$}
\address{$^1$SEECS, University of Wales, Bangor LL57 1UT, UK}
\address{$^2$Hewlett-Packard Labs, Mail Box M48, Bristol BS34 8QZ, UK}

\maketitle

\begin{abstract} 
 We study the experimental realisation of quantum teleportation as performed
 by Bouwmeester {\em et al}.\ [Nature {\bf 390}, 575 (1997)] and the 
 adjustments to it suggested by Braunstein and Kimble [Nature {\bf 394}, 841
 (1998)]. These suggestions include the employment of a detector cascade and a 
 relative slow-down of one of the two down-converters. We show that 
 coincidences between photon-pairs from parametric down-conversion
 automatically probe the non-Poissonian structure of these sources. 
 Furthermore, we find that detector cascading is of limited use, and that
 modifying the relative strengths of the down-conversion efficiencies will
 increase the time of the experiment to the order of weeks. Our analysis
 therefore points to the benefits of single-photon detectors in 
 non-post-selected type experiments, a technology currently requiring roughly
 $6^{\circ}$K operating conditions.
\end{abstract}

PACS number(s): 03.67.*, 42.50.Dv

\medskip

\begin{multicols}{2}

Quantum entanglement, an aspect of quantum theory already recognised in the 
early days, clearly sets quantum mechanics apart from classical mechanics. 
More recently, fundamentally new phenomena involving entanglement such as 
cryptography, error correction and dense coding have been discovered 
\cite{crypto,errorcor,dense}. In particular, the field has witnessed major 
steps forward with the experimental realisation of {\em quantum teleportation} 
\cite{bennett,bouwmeester,boschi,nmr,kimble,science}.

We speak of quantum teleportation when a (possibly unknown) quantum state 
$|\phi\rangle$ held by Alice is sent to Bob without actually traversing the 
intermediate space. The protocol uses an entangled state of two systems which 
is shared between Alice and Bob. To bring teleportation about, Alice and Bob 
proceed as follows: a Bell measurement of $|\phi\rangle$ and Alice's half of 
the entangled pair will correlate Bob's half to the original state 
$|\phi\rangle$. Bob then uses the outcome of Alice's measurement to determine 
which unitary transformation brings his state into the original one
$|\phi\rangle$. 

In this paper, we study the experimental realisation of quantum teleportation 
of a single polarised photon as performed in Innsbruck, henceforth called the 
`Innsbruck experiment' (Bouwmeester {\em et al}.\ \cite{bouwmeester}). Our aim 
is to evaluate the suggestions to `improve' the experiment in order to yield 
non-post-selected operation, as given by Braunstein and Kimble 
\cite{braunstein,comment}. These suggestions include the employment of a 
so-called detector cascade in the state preparation mode, and enhancement of 
the photon-pair source responsible for the entanglement channel relative to 
the one responsible for the initial state preparation. 

Subsequently, we hope to clarify some of the differences in the interpretation 
of the Innsbruck experiment. As pointed out by Braunstein and Kimble 
\cite{braunstein}, to lowest order the teleported state in the Innsbruck 
experiment is a mixture of the vacuum and a single-photon state. However, we 
cannot interpret this {\em state} as a low-efficiency teleported state, where 
sometimes a photon emerges from the apparatus and sometimes not. This reasoning
is based on what we call the `Partition Ensemble Fallacy', or {\sc pef} for 
short. It relies on a particular partition of the outgoing density matrix, and 
this is not consistent with quantum mechanics \cite{peres}. Circumventing {\sc
pef} leads to the notion of {\em post-selected} teleportation, in which the
teleported state is detected. The post-selected teleportation indeed has a high
fidelity and a low efficiency. Although generally {\sc pef} is harmless (it 
might even be considered a useful tool in understanding aspects of quantum 
theory), to our knowledge, this is the first instance where it leads to a {\em 
quantitatively } different evaluation of an experiment.

The main result of this paper is that the suggested improvements require
near perfect efficiency photo-detectors or a considerable increase in the time
needed to run the experiment. The remaining practical alternative in order to 
obtain {\em non-post-selected} quantum teleportation (i.e., teleportation {\em 
without} the need for detecting the teleported photon) is to employ a
single-photon detector in the state-preparation mode (a technology currently 
requiring approximately $6^{\circ}$K operating conditions).

We start in the next section by reviewing photon-pair creation using
parametric down-conversion. In section \ref{whatisqt} we present the fidelity 
of teleportation and discuss some of its interpretations. Finally, section 
\ref{generalised} is devoted to an analysis of a generalised version of the 
Innsbruck experiment and requirements are given which sufficiently enhance the 
fidelity.

\section{The Innsbruck Experiment}\label{experiment}

In this section we review the Innsbruck experiment. In section \ref{sec:npairs}
we calculate the probability distribution of finding $n$ photon-pairs and 
subsequently we compare this with the Poisson distribution for $n$ 
photon-pairs. The difference between the two distributions, in terms of 
distinguishability, is evaluated by means of the so-called {\em statistical 
distance} in section \ref{stat}.

In the Innsbruck experiment, parametric down-conversion is used to create two 
entangled photon-pairs. One pair constitutes the entangled state shared 
between Alice and Bob, while the other is used by Victor to create an `unknown'
single-photon polarisation state $|\phi\rangle$: Victor detects mode $a$, shown
in figure \ref{fig:innsbruck} to prepare the single-photon input state in mode 
$b$. This mode is sent to Alice. A coincidence in the detection of the two 
outgoing modes of the beam-splitter (Alice's --- incomplete --- Bell 
measurement) tells us that Alice's two photons are in a $|\Psi^-\rangle$ Bell 
state \cite{weinfurter,mann}. The remaining photon (held by Bob) is now in the 
same unknown state as the photon prepared by Victor because in this case the 
unitary transformation Bob has to apply coincides with the identity, i.e., 
doing nothing. Bob verifies this by detecting his state along the same 
polarisation axis which was used by Victor. A four-fold coincidence in the 
detectors of Victor's state preparation, Alice's Bell measurement and Bob's 
outgoing state indicate that quantum teleportation of a single-photon state 
is complete.

\figone

There is however a complication which gave rise to a different interpretation
of the experiment \cite{braunstein,comment}. Analysis shows that the state 
detected by Bob is a mixture of the vacuum and the original state 
\cite{bouwmeester,braunstein} (to lowest order). This vacuum contribution 
occurs when the down-converter responsible for creating the input state 
$|\phi\rangle$ yields two photon-pairs, while the other gives nothing. The 
detectors used in the experiment cannot distinguish between one or several 
photons coming in, so Victor's detection of mode $a$ in figure 
\ref{fig:innsbruck} will not reveal the presence of more than one photon. A 
three-fold coincidence in the detectors of Victor and Alice is still possible, 
but Bob has not received a photon and quantum teleportation has not been 
achieved. Bob therefore needs to detect his state in order to identify 
successful quantum teleportation. When Victor uses a detector which can 
distinguish between one or several photons this problem vanishes. However, 
currently such detectors require an operating environment of roughly 
$6^{\circ}$K \cite{kwiat1994,hughes,kim}.

In section \ref{generalised} we give a detailed analysis of the Innsbruck 
experiment and the suggestions for improvement given in Ref.\ \cite{comment}. 
Here, we investigate the creation of entangled photon-pairs using weak 
parametric down-conversion \cite{kwiat}. In this process, there is a small 
probability of creating more than one photon-pair simultaneously. One might 
expect that for sufficiently weak down-conversion the two pairs created by one 
source (which give rise to the vacuum contribution in the teleported output 
state) can be considered independent from each other. However, we show that 
this is not the case. In what follows we find it convenient to `unfold' the 
experimental setup according to figure \ref{model}.

\figtwo

\subsection{Probability for $n$ pairs}\label{sec:npairs}

In this section we study the statistics of parametric down-conversion. We 
show that the probability $P_{\text{PDC}}(n)$ for finding $n$ photon-pairs
deviates from the Poisson distribution, even in the weak limit.

Let $a$ and $b$ be two field modes with a particular polarisation along the 
$x$- and $y$-axis of a given coordinate system. We are working in the 
interaction picture of the Hamiltonian which governs the dynamics of creating 
two entangled field modes $a$ and $b$ using weak parametric down-conversion. 
In the rotating wave approximation this Hamiltonian reads ($\hbar =1$):
\begin{equation}\label{hamilton}
  H = i\kappa ( a_x^{\dagger} b_y^{\dagger} -
    a_y^{\dagger} b_x^{\dagger} ) + \text{H.c.}
\end{equation}
In this equation H.c.\ means Hermitian conjugate, and $\kappa$ is the product
of the pump amplitude and the coupling constant between the {\sc em}-field and 
the crystal. The operators $a^{\dagger}_i$, $b^{\dagger}_i$ and $a_i$, $b_i$ 
are creation and annihilation operators for polarisations $i\in\{ x,y \}$ 
respectively. They satisfy the 
following commutation relations:
\begin{eqnarray}\label{commutation}
 [a_i,a_j^{\dagger}] = \delta_{ij}~ , &\qquad & 
	[a_i,a_j] = [a_i^{\dagger},a_j^{\dagger}] = 0, \cr
 [b_i,b_j^{\dagger}] = \delta_{ij}~ , &\qquad &
	[b_i,b_j] = [b_i^{\dagger},b_j^{\dagger}] = 0 \; ,
\end{eqnarray}
where $i,j\in\{ x,y \}$. The time evolution due to 
this Hamiltonian is given by
\begin{equation}\label{unitary}
 U(t)\equiv\exp(-iHt) \; ,
\end{equation}
where $t$ is the time it takes for the pulse to travel through the crystal.
By applying this unitary transformation to the vacuum $|0\rangle$ the state 
$|\Psi_{\text{src}}\rangle$ is obtained:
\begin{equation}\label{evolution}
  |\Psi_{\text{src}}\rangle = U(t) |0\rangle = \exp(-iHt) |0\rangle \; .
\end{equation}

We are interested in the properties of $|\Psi_{\text{src}}\rangle$. Define the 
$L_+$- and the $L_-$-operator to be
\begin{equation}
 L_+ = a_x^{\dagger} b_y^{\dagger}-a_y^{\dagger} b_x^{\dagger}= L_-^{\dagger}
 \; .
\end{equation}
This will render Eqs.\ (\ref{hamilton}) and (\ref{unitary}) into:
\begin{eqnarray}\label{unitl+}
 {H} &=& i\kappa L_+ - i\kappa^* L_- \qquad\text{and}\cr
        U(t) &=& \exp[\kappa t L_+ - \kappa^* t L_-] \; .
\end{eqnarray}

Applying $L_+$ to the vacuum will yield a singlet state (up to a normalisation 
factor) in modes $a$ and $b$:
\begin{eqnarray}
  L_+ |0\rangle &=& |\leftrightarrow,\updownarrow\rangle_{ab} - 
		|\updownarrow,\leftrightarrow\rangle_{ab} \cr
  &=& |1,0;0,1\rangle_{a_x a_y b_x b_y}-|0,1;1,0\rangle_{a_x a_y b_x b_y} \; ,
\end{eqnarray}
we henceforth use the latter notation where $|i,j;k,l\rangle_{a_x a_y b_x b_y}$
is shorthand for $|i\rangle_{a_x}\!\otimes|j\rangle_{a_y}\!\otimes|k
\rangle_{b_x}\!\otimes|l\rangle_{b_y}$, a tensor product of photon number 
states. Applying this operator $n$ times gives a state $|\Phi^n\rangle$ (where 
we have included a normalisation factor $N_n$, so that 
$\langle\Phi^n|\Phi^n\rangle=1$):
\begin{eqnarray}
  |\Phi^n\rangle &\equiv& N_n L_+^n |0\rangle = N_n \sum_{m=0}^{n} n! (-1)^m\cr
  && \qquad |m_x,(n-m)_y;(n-m)_x,m_y \rangle_{ab} \; ,
\end{eqnarray}
with 
\begin{equation}\label{normalization}
  N_n^2 = \frac{1}{n!(n+1)!} \; .
\end{equation}
We interpret $|\Phi^n\rangle$ as the state of $n$ entangled photon-pairs.

We want the unitary operator $U(t)$ in Eq.\ (\ref{unitl+}) to be in a 
normal ordered form, because then the annihilation operators will `act' on 
the vacuum first, in which case Eq.\ (\ref{evolution}) simplifies. In order to 
obtain the normal ordered form of $U(t)$ we examine the properties of $L_+$ 
and $L_-$. Given the commutation relations (\ref{commutation}), it is 
straightforward to show that:
\begin{eqnarray}
  [L_-,L_+] &=& a_x^{\dagger} a_x + a_y^{\dagger} a_y + b_x^{\dagger} b_x + 
	b_y^{\dagger} b_y + 2 \cr &\equiv&  2 L_0 \quad\text{and}\cr
  [L_0,L_{\pm}] &=& \pm L_{\pm} \; .
\end{eqnarray}
An algebra which satisfies these commutation relations (together with the 
properties $L_- = L_+^{\dagger}$ and $L_0 = L_0^{\dagger}$) is an $su(1,1)$ 
algebra. The normal ordering for this algebra is known \cite{truax} (with 
$\hat\tau = \tau/|\tau|$):
\begin{eqnarray}\label{BCHSU11}
 \exp\left(\tau L_+ - \tau^* L_-\right) &=&
	\exp\left(\hat\tau\tanh|\tau| L_+\right) \cr
	&& \quad\times \exp\left[-2\ln(\cosh|\tau|) L_0\right] \cr
	&& \quad\times \exp\left(-\hat\tau^*\tanh|\tau|L_-\right) \; . 
\end{eqnarray}
The scaled time $\tau$ is defined as $\tau\equiv\kappa t$. Without loss of 
generality we can take $\tau$ to be real. Since the `lowering' operator $L_-$
is placed on the right, it will yield zero when applied to the vacuum and 
the exponential reduces to the identity. Similarly, the exponential containing 
$L_0$ will yield a $c$-number, contributing only to the normalisation.

We can now ask the question whether the pairs thus formed are independent of 
each other, i.e., whether they yield the Poisson distribution. Suppose 
$P_{\text{PDC}}(n)$ is the probability of creating $n$ photon-pairs with 
parametric down-conversion and let 
\begin{equation}\label{substitutions}
  r \equiv \tanh\tau\qquad\text{and}\qquad
  q \equiv 2\ln(\cosh\tau) \; , 
\end{equation}
then the probability of finding $n$ entangled photon-pairs is:
\begin{eqnarray}\label{EPRdist}
  P_{\text{PDC}}(n) &\equiv& |\langle \Phi^n |\Psi_{\text{src}}\rangle|^2 \cr 
  &=& |\langle 0 | \left( L_-^n N_n \right) \left( e^{rL_+}e^{-qL_0} e^{-rL_-} 
      \right) |0\rangle|^2 \cr
  &=& e^{-2q} |\langle 0 | L_-^n N_n \left[ \sum_{l=0}^{\infty} \frac{r^l}{l!} 
  L_+^l\right] |0\rangle|^2 \cr 
  &=& (n+1) r^{2n} e^{-2q} \; .
\end{eqnarray}
It should be noted that this is only a normalised probability distribution in 
the limit of $r,q\rightarrow 0$. 

Given Eqs.\ (\ref{substitutions}) $P_{\text{PDC}}(n)$ deviates from the 
Poisson distribution, and the pairs are therefore not independent. For weak 
sources, however, one might expect that $P_{\text{PDC}}(n)$ approaches the 
Poisson distribution sufficiently closely. This hypothesis can be tested by 
studying the distinguishability of the two distributions.

\subsection{Distinguishability}\label{stat}

Here, we study the distinguishability of the pair distribution calculated in 
the previous section and the Poisson distribution. The Poisson distribution for
independently created objects is given by
\begin{equation}\label{poisson}
 P_{\text{poisson}}(n) = \frac{p^n e^{-p}}{n!} \; .
\end{equation}
Furthermore, rewrite the pair distribution in Eq.\ (\ref{EPRdist}) as
\begin{equation}\label{EPRdist2}
  P_{\text{PDC}}(n) = (n+1) \left(\frac{p}{2}\right)^n e^{-p} \qquad
  \text{for}~p\ll 1 \; ,
\end{equation}
using $q\approx r^2$ and $p \equiv 2r^2 = 2\tanh^2\tau$ for small scaled times.
Here $p$ is the probability of creating one entangled photon-pair. Are these 
probability distributions distinguishable? Naively one would say that for 
sufficiently weak down-conversion (i.e., when $p\ll 1$) these distributions 
largely coincide, so that instead of the complicated pair-distribution 
(\ref{EPRdist2}) we can use the Poisson distribution, which is much easier 
from a mathematical point of view. The distributions are distinguishable when 
the `difference' between them is larger than the size of an average statistical
fluctuation of the difference. This fluctuation depends on the number of 
samplings. 

Consider two nearby discrete probability distributions $\{ p_j\}$ and $\{ p_j 
+ \text{d}p_j\}$. A natural difference between these distributions is given by 
the so-called (infinitesimal) {\em statistical distance} d$s$ 
\cite{wootters,braun1994,jos}:
\begin{equation}
 \text{d}s^2 = \sum_j \frac{\text{d}p_j^2}{p_j} \; .
\end{equation}
When the typical statistical fluctuation after $N$ samplings is 
$1/\sqrt{N}$, the two probability distributions are distinguishable if:
\begin{equation}
 \text{d}s \gtrsim \frac{1}{\sqrt{N}} \quad\Leftrightarrow\quad 
	N\text{d}s^2 \gtrsim 1 \; .
\end{equation}
The statistical distance between (\ref{poisson}) and (\ref{EPRdist2}), and 
therefore the distinguishability criterion is: 
\begin{equation}\label{32p2}
 \text{d}s^2 \propto \frac{p^2}{8} \quad\rightarrow\quad N \gtrsim 
	\frac{8}{p^2} \; .
\end{equation}
On the other hand, the average number of trials in the teleportation 
experiment required to get one photon-pair from both down-converters is:
\begin{equation}\label{speed}
 N = \frac{1}{p^2} \; .
\end{equation}
The minimum number of trials in the experiment thus almost immediately renders 
the two probability distributions distinguishable, and we therefore cannot 
approximate the actual probability distribution with the Poisson distribution.

Since the Poisson distribution in Eq.\ (\ref{poisson}) is derived by requiring 
statistical independence of $n$ pairs and the pair distribution is 
distinguishable from the Poisson distribution, the photon-pairs cannot be 
considered to be independently produced, {\em even in the weak limit}. In the 
analysis of the Innsbruck experiment we need to take extra care due to this 
property of parametric down-converters.

\section{Teleportation Fidelity}\label{whatisqt}

In this section we introduce the so-called {\em fidelity} for quantum 
teleportation. This is already recognised as an important tool in quantum 
information theory and it is therefore natural to consider teleportation 
criteria based upon it. Subsequently, we discuss different points of view of 
the Innsbruck experiment emerging from this concept. We restrict our discussion
to the subset of events where successful Bell-state and state-preparation 
detections have occurred (all subsequent statements are {\em conditioned} on 
such events). Since the interpretation of the experiment has become a slightly 
controversial issue, we treat this in some detail. 

In order to define the fidelity, denote the input state by $|\phi\rangle$ 
(which is here assumed to be pure) and the outgoing (teleported) state by a 
density matrix $\rho_{\text{out}}$. The fidelity $F$ is the overlap between 
the incoming and the outgoing state:
\begin{equation}\label{fiddef}
 F = \text{Tr}[\rho_{\text{out}}|\phi\rangle\langle\phi|] \; .
\end{equation}
It corresponds to the lower bound for the probability of mistaking 
$\rho_{\text{out}}$ for $|\phi\rangle$ in any possible (single) measurement
\cite{fuchs}. When $\rho_{\text{out}}$ is an exact replica of $|\phi\rangle$ 
then $F=1$, and when $\rho_{\text{out}}$ is a imprecise copy of $|\phi\rangle$ 
then $F<1$. Finally, when $\rho_{\text{out}}$ is completely othogonal to 
$|\phi\rangle$ the fidelity is zero. 

In the context of this paper, the fidelity is used to distinguish between
quantum teleportation and teleportation which could have been achieved 
`classically'. Classical teleportation is the disembodied transport of some
quantum state from Alice to Bob by means of a classical communication channel.
There is {\em no} shared entanglement between Alice and Bob. Since classical 
communication can be duplicated, such a scheme can lead to many copies of the
transported output state (so-called {\em clones}). Classical teleportation 
with perfect fidelity (i.e., $F=1$) would then lead to the possibility of 
{\em perfect} cloning, thus violating the no-cloning theorem 
\cite{wootters1982}. This means that the maximum fidelity for classical 
teleportation has an upper bound which is less than one. 

Quantum teleportation, on the other hand, can achieve perfect fidelity (and 
circumvents the no-cloning theorem by disrupting the original). To demonstrate 
{\em quantum} teleportation therefore means \cite{note} that the teleported 
state should have a higher fidelity than possible for a state obtained by any 
scheme involving classical communication {\em alone}.

For classical teleportation of randomly sampled polarisations, the maximum 
attainable fidelity is $F=2/3$. When only linear polarisations are to be
teleported, the maximum attainable fidelity is $F=3/4$ 
\cite{fuchs1996,massar1999,massar,fuchs}. These are the values which the 
quantum teleportation fidelity should exceed.

In the case of the Innsbruck experiment, $|\phi\rangle$ denotes the `unknown' 
linear polarisation state of the photon issued by Victor. We can write the 
undetected outgoing state to lowest order as
\begin{equation}\label{rhoout}
 \rho_{\text{out}} \propto |\alpha|^2 |0\rangle\langle 0| + |\beta|^2 
   |\phi\rangle\langle\phi| \; ,
\end{equation}
where $|0\rangle$ is the vacuum state. The overlap between $|\phi\rangle$ and 
$\rho_{\text{out}}$ is given by Eq.\ (\ref{fiddef}). In the Innsbruck 
experiment the fidelity $F$ is then given by
\begin{equation}\label{fidelity}
 F \equiv \text{Tr}[\rho_{\text{out}}|\phi\rangle\langle\phi|] = 
 \frac{|\beta|^2}{|\alpha|^2 + |\beta|^2}\; .
\end{equation}
This should be larger than 3/4 in order to demonstrate quantum teleportation. 
The vacuum contribution in Eq.\ (\ref{rhoout}) arises from the fact that 
Victor cannot distinguish between one or several photons entering his detector,
i.e., Victor's inability to properly prepare a single-photon state.

As pointed out by Braunstein and Kimble \cite{braunstein}, the fidelity of the
Innsbruck experiment remains well below the lower bound of 3/4 due to the 
vacuum contribution (the exact value of $F$ will be calculated in the next
section). Replying to this, Bouwmeester {\em et al}.\ \cite{comment,jmod} 
argued that `when a photon appears, it has all the properties required by the 
teleportation protocol'. The vacuum contribution in Eq.\ (\ref{rhoout}) should 
therefore only affect the efficiency of the experiment, with a consequently 
high fidelity. However, this is a potentially ambiguous statement. If by 
`appear' we mean `appearing in a photo-detector', we agree that a high 
fidelity (and low efficiency) can be inferred. However, this yields a 
so-called {\em post-selected} fidelity, where the detection destroys the 
teleported state. The fidelity prior to (or without) Bob's detection is called
the {\em non}-post-selected fidelity. The question is now whether we can say 
that a photon appears when {\em no} detection is made, thus yielding a high
{\em non}-post-selected fidelity. 

This turns out not to be the case. Making an {\em ontological} distinction 
between a photon and {\em no} photon in a mixed state (without a detection) is 
based on what we call the `Partition Ensemble Fallacy.' We now study this in 
more detail.

Consider the state $\rho_{\rm out}$ of the form of Eq.\ (\ref{rhoout}). To 
lowest order, it is the sum of two pure states. However, this is not a unique 
`partition'. Whereas in a chemical mixture of, say, nitrogen and oxygen there 
is a unique partition (into N$_2$ and O$_2$), a quantum mixture can be 
decomposed many ways. For instance, $\rho_{\rm out}$ can equally be written in 
terms of
\begin{equation}
 |\psi_1\rangle = \alpha |0\rangle + \beta |\phi\rangle \quad\text{and}\quad
 |\psi_2\rangle = \alpha |0\rangle - \beta |\phi\rangle
\end{equation}
as
\begin{equation}\label{novacuum}
 \rho_{\rm out} = \frac{1}{2} |\psi_1\rangle\langle\psi_1| + \frac{1}{2} 
	|\psi_2\rangle\langle\psi_2|\; .
\end{equation}
In fact, this is just one of an infinite number of possible decompositions. 
Quantum mechanics dictates that all partitions are equivalent to each other 
\cite{peres}. They are indistinguishable. To elevate one partition over another
is to commit the `Partition Ensemble Fallacy.'

Returning to the Innsbruck experiment, we observe that in the absence
of Bob's detection, the density matrix of the teleported state 
(i.e., the {\it non}-post-selected state) may be decomposed into an infinite 
number of partitions. These partitions do not necessarily include the 
vacuum state at all, as exemplified in Eq.\ (\ref{novacuum}). It would 
therefore be incorrect to say that teleportation did or did not occur except
through some operational means (e.g., a detection performed by Bob).

Bob's detection thus leads to a high post-selected fidelity. However, the
vacuum term in Eq.\ (\ref{rhoout}) contributes to the {\em non}-post-selected
fidelity, decreasing it well below the lower bound of 3/4 (see the next 
section). Due to this vacuum contribution, the Innsbruck experiment did {\em 
not} demonstrate {\em non}-post-selected quantum teleportation. Nonetheless, 
teleportation was demonstrated using post-selected data obtained by detecting 
the teleported state. By selecting events where a photon was observed in the 
teleported state, a post-selected fidelity higher than $3/4$ could be inferred 
(estimated at roughly 80\% \cite{jmod}). [We recall that this entire 
discussion is restricted to the subset of events where successful Bell-state 
and state-preparation have occurred.]

\section{Generalised Experiment}\label{generalised}

In this section we present a generalised scheme for the Innsbruck experiment 
which enables us to establish the requirements to obtain non-post-selected 
quantum teleportation (based on a three-fold coincidence of Victor and Alice's 
detectors). The generalisation consists of a detector cascade \cite{song} for 
Victor's state-preparation detection and parametric down-converters with 
different specifications, rather than two identical down-converters. We 
consider a detector cascade since single-photon detectors currently require
roughly $6^{\circ}$K operating conditions \cite{kim}. Furthermore, an 
arbitrary polarisation rotation in the state-preparation mode allows us to 
consider any superposition of $x$- and $y$-polarisation.

First, we give an expression for detectors with a finite efficiency. Then we 
calculate the output state and give an expression for the teleportation 
fidelity in terms of the detector efficiencies and down-converter 
probabilities. 

\subsection{Detectors}\label{detectors}

There are two sources of errors for a detector: it might fail to detect a 
photon, or it might give a signal although there wasn't actually a photon 
present. The former is called a `detector loss' and the latter a `dark count'. 
Dark counts are negligible in the teleportation experiment because the {\sc 
uv}-pump is fired during very short time intervals and the probability of 
finding a dark count in such a small interval is negligible. Consequently, the 
model for real, finite-efficiency detectors we adopt here only takes into 
account detector losses. Furthermore, the detectors cannot distinguish between 
one or several photons.

To simulate a realistic detector we make use of projection operator valued 
measures, or {\sc povm}'s for short \cite{kraus}. Consider a beam-splitter 
in the mode which is to be detected so that part of the signal is reflected 
(see figure \ref{fig:detector}). The second incoming mode of the beam-splitter 
is the vacuum (we neglect higher photon number states because they hardly 
contribute at room temperature). The transmitted signal $c$ is sent into an 
ideal detector. We identify mode $d$ with the detector loss.

\figthree

Suppose in mode $a$ there are $n$ $x$-polarised and $m$ $y$-polarised photons.
Furthermore, let these photons all be reflected by the beam-splitter (since the
detectors cannot distinguish between one or more photons, we do not consider 
the case where only some of the photons are reflected; we are interested in a 
`click' in the detector and partially reflected modes still give a click). The 
projector for finding these photons in the $d$-mode is given by:
\begin{eqnarray}\label{povmnm}
  {E}_d &=& |n,m\rangle_{d_x d_y}\langle n,m| \cr &=& \frac{1}{n!m!}
  (d_x^{\dagger})^n(d^{\dagger}_y)^m |0,0\rangle_{d_x d_y}\langle 0,0| 
	d^n_x d^m_y \; .
\end{eqnarray}
The beam-splitter equations are taken to be ($\widetilde\eta\equiv
\sqrt{1-\eta^2}$):
\begin{equation}\label{bs}
  c = \eta a + \widetilde\eta b \qquad\text{and}\qquad
  d = \widetilde\eta a - \eta b \; .
\end{equation}
Substituting these equations in (\ref{povmnm}), summing over all $n$ and 
$m$ and using the binomial expansion yields
\end{multicols}

\noindent\rule{5cm}{.5pt}

\begin{eqnarray}
  {E}_{ab} &=& \sum_{n,m} \binom{n}{k}^2 \binom{m}{l}^2 
	\frac{(-1)^{2(k+l)}}{n!m!}
	(\widetilde\eta a^{\dagger}_x)^{n-k} (\eta b^{\dagger}_x)^k 
	(\widetilde\eta a^{\dagger}_y)^{m-l} (\eta b^{\dagger}_y)^l~ 
	|0\rangle_{ab} \cr && \times \langle 0| (\widetilde\eta a_x)^{n-k} 
	(\eta b_x)^k (\widetilde\eta a_y)^{m-l} (\eta b_y)^l \; .
\end{eqnarray}
Since the $b$-mode is the vacuum, the only contributing term is $k=l=0$. So 
the {\sc povm} ${E}_a^{(0)}$ of finding {\em no} detector counts in mode 
$a$ is
\begin{equation}
 {E}_a^{(0)} = \sum_{n,m} \frac{ \widetilde{\eta}^n 
  (a_x^{\dagger})^n \widetilde{\eta}^m (a_y^{\dagger})^m}{n!m!} 
	~|0\rangle_{a_x a_y}\langle 0|~
  \widetilde{\eta}^n a_x^n \widetilde{\eta}^m a_y^m
  = \sum_{n,m} \widetilde\eta^{2(n+m)} |n,m\rangle_{a_x a_y} \langle n,m| \; .
\end{equation}
The required {\sc povm} for finding a detector count is 
\begin{equation}
  {E}_a^{(1)} = I - {E}_a^{(0)} = \sum_{n,m} 
  [1-\widetilde\eta^{2(n+m)}] |n,m\rangle_{a_x a_y} \langle n,m| \; ,
\end{equation}

\hfill\rule{5cm}{.5pt}

\begin{multicols}{2}
where $I$ is the unity operator, $\eta^2$ is the detector efficiency and
$\widetilde\eta^2$ the detector loss. 
When we let ${E}_a^{(1)}$ act on the total state and trace out mode 
$a$, we have inefficiently detected this mode. However, it is worth noting 
that this model only applies for short periods of detection. In the case of 
continuous detection we need a more elaborate model (see e.g.\ Ref.\ 
\cite{wiseman}).

In order for Victor to distinguish between one or more photons in the state
preparation mode $a$, we consider a detector cascade (Victor doesn't have a 
detector which can distinguish between one or several photons coming in). When 
there is a detector coincidence in the cascade, more than one photon was 
present in mode $a$, and the event should be dismissed. In the case of ideal 
detectors, this will improve the fidelity of the teleportation up to an 
arbitrary level (we assume there are no beam-splitter losses). Since we employ 
the cascade in the $a$-mode (which was used by Victor to project mode $b$ onto 
a superposition in the polarisation basis) we need to perform a {\em 
polarisation sensitive} detection.

In order to model this we separate the incoming state $|n,m\rangle_{a_x a_y}$ 
of mode $a$ into two spatially separated modes $|n\rangle_{a_x}$ and 
$|m\rangle_{a_y}$ by means of a polarisation beam-splitter (see figure 
\ref{fig:cascade}). The modes $a_x$ and $a_y$ will now be detected. The {\sc 
povm}'s corresponding to inefficient detectors are derived along the same 
lines as in the previous section and read:
\begin{eqnarray}\label{cascadepovm}
 {E}^{(0)}_{a_j} &=& \sum_n\widetilde{\eta}^{2n}|n\rangle_{a_j}\langle 
	n| \qquad\text{and}\cr
 {E}^{(1)}_{a_j} &=& \sum_n[1-\widetilde{\eta}^{2n}]|n\rangle_{a_j}
	\langle n| \; .
\end{eqnarray}
with $j\in\{ x,y \}$. We choose to detect the $x$-polarised mode. This means
that we only have to make sure that there are no photons in the $y$-mode.
The output state will include a product of the two {\sc povm}'s: one for
finding a photon in mode $a_x$, and one for finding {\em no} photons in mode
$a_y$: $E^{(1)}_{a_x}E^{(0)}_{a_y}$.

\figfour

To make a cascade with two detectors in $a_x$ and one in $a_y$ employ another 
50:50 beam-splitter in mode $a_x$ and repeat the above procedure of detecting 
the outgoing modes $c$ and $d$ (\ref{cascadepovm}). Since we can detect a 
photon in either one of the modes, we have to include the sum of the 
corresponding {\sc povm}'s, yielding a transformation $E^{(1)}_{c_x} 
E^{(0)}_{d_x} + E^{(0)}_{c_x}E^{(1)}_{d_x}$. This is easily expandable to 
larger cascades by using more beam-splitters and summing over all possible 
detector hits.

\subsection{Output state}

In this section we incorporate the finite-efficiency detectors and the detector
cascade in our calculation of the undetected teleported output state. This 
calculation includes the creation of two photon-pairs (lowest order) and three 
photon pairs (higher order corrections due to four or more photon-pairs in the 
experiment are highly negligible). A formula for the vacuum contribution to 
the teleportation fidelity is given for double-pair production (lowest order). 

Let the two down-converters in the generalised experimental setup yield 
evolutions $U_{\text{src1}}$ and $U_{\text{src2}}$ on modes $a$, $b$ and $c$,
$d$ respectively (see figures \ref{fig:innsbruck} and \ref{model}) according 
to (\ref{evolution}). The beam-splitter which transforms modes $b$ and $c$ 
into $u$ and $v$ (see figure \ref{model}) is incorporated by a suitable 
unitary transformation $U_{\text{BS}}$, as is the polarisation rotation 
$U_{\theta}$ over an angle $\theta$ in mode $a$. The $n$-cascade will be 
modelled by $n-1$ beam-splitters in the $x$-polarisation branch of the cascade,
and can therefore be expressed in terms of a unitary transformation 
$U_{a_1\ldots a_n}$ on the Hilbert space corresponding to modes $a_1$ to $a_n$ 
(i.e., replace mode $a$ with modes $a_1$ to $a_n$):
\begin{eqnarray}\label{subout}
 |\Psi_{\theta}\rangle\langle\Psi_{\theta}|
  &=& U_{a_1\ldots a_n} U_{\theta} U_{\text{BS}}U_{\text{scr1}}U_{\text{scr2}}
    |0\rangle \cr 
  &&\qquad\times\langle 0| U^{\dagger}_{\text{scr1}} U^{\dagger}_{\text{scr2}} 
    U^{\dagger}_{\text{BS}} U^{\dagger}_{\theta} U^{\dagger}_{a_1\ldots a_n}\; 
    .
\end{eqnarray}
Detecting modes $a_1\ldots a_n$, $u$ and $v$ with real (inefficient) detectors 
means taking the partial trace over the detected modes, including the {\sc 
povm}'s derived in section \ref{detectors}:
\begin{equation}\label{out}
 \rho_{\text{out}} = \text{Tr}_{a_1\ldots a_n u v} \left[ 
 {E}_{n\text{-cas}} {E}^{(1)}_u {E}^{(1)}_v 
 ~|\widetilde{\Psi}_{\theta}\rangle_{a_1\ldots a_n uvd}\langle
 \widetilde{\Psi}_{\theta}| \right] \; ,
\end{equation}
with ${E}_{n\text{-cas}}$ the superposition of {\sc povm}'s for a 
polarisation sensitive detector cascade having $n$ detectors with finite 
efficiency. In the case $n=2$ this expression reduces to the 2-cascade {\sc 
povm}-superposition derived in the previous section. Eq.\ (\ref{out}) is an 
analytic expression of the undetected outgoing state in the generalisation of 
the Innsbruck experiment.

The evolutions $U_{\text{src1}}$ and $U_{\text{src2}}$ are exponentials of 
creation operators. In the computer simulation (using Mathematica) we 
truncated these exponentials at first and second order. The terms that remain 
correspond to double and triple pair production in the experimental setup. To 
preserve the order of the creation operators we put them as arguments in a 
function $f$. We defined the following algebraic rules for $f$:
\begin{eqnarray}
  f[x, y+w, z] &:=& f[x,y,z] + f[x,w,z]\; , \cr
  f[x, n a, y] &:=& n f[x,a ,y] \; , \cr
  f[x, n a^{\dagger}, y] &:=& n f[x,a^{\dagger} ,y] \; ,
\end{eqnarray}
where $x,y,z$ and $w$ are arbitrary expressions including creation and 
annihilation operators ($a^{\dagger}$ and $a$) and $n$ some expression {\em 
not} depending on creation or annihilation operators. The last entry of $f$ is 
always a photon number state (including the initial vacuum state).

Since we now have functions of creation and annihilation operators, it is 
quite straightforward to define (lists of) substitution rules for a
beam-splitter (see also Eq.\ (\ref{bs})), polarisation rotation, {\sc povm}'s 
and the trace operation. We then use these substitution rules
to `build' a model of the generalised experimental setup.

\subsection{Results}

The probability of creating one entangled photon-pair using the weak 
parametric down-conversion source 1 or 2 is $p_1$ or $p_2$ respectively (see 
figure \ref{model}). We calculated the output state both for an $n$-cascade up 
to order $p^2$ (i.e.\ $p_1^2$ or $p_1 p_2$) and for a 1-cascade up to the order
$p^3$ ($p_1^3$, $p_1^2 p_2$ or $p_1 p_2^2$). The results are given below. For 
brevity, we take:
\begin{eqnarray}
 |\Psi_\theta\rangle &=& \cos\theta |0,1\rangle + \sin\theta |1,0\rangle
	\qquad\text{and}\cr
 |\Psi^{\perp}_\theta\rangle &=& \sin\theta |0,1\rangle -\cos\theta|1,0\rangle
\end{eqnarray}
as the ideally prepared state and the state orthogonal to it. Suppose 
$\eta^2_u$ and $\eta^2_v$ are the efficiencies of the detectors in mode $u$ 
and $v$ respectively, and $\eta^2_c$ the efficiency of the detectors in the 
cascade (for simplicity we assume that the detectors in the cascade have the 
same efficiency). Define $g_{uvc} = \eta^2_u \eta^2_v \eta^2_c$. The detectors 
in modes $u$ and $v$ are polarisation insensitive, whereas the cascade 
consists of polarisation sensitive detectors. Bearing this in mind, we have up 
to order $p^2$ for an $n$-cascade in mode $a_x$ and finding no detector click 
in the $a_y$-mode:
\begin{eqnarray}\label{ord2cas2}
 \rho_{\text{out}} \propto \frac{p_1}{8}g_{uvc} \biggl\{ && \frac{p_1}{n} 
  [1+(5n-3)(1-\eta^2_c)]|0\rangle\langle 0|\cr && ~+ p_2 
  |\Psi_\theta\rangle\langle\Psi_\theta|\biggr\} + O(p^3) \; ,
\end{eqnarray}
where the vacuum contribution formula was calculated and found to be correct 
for $n\leq 4$ (and $n\neq 0$). 

In order to have non-post-selected quantum teleportation, the fidelity $F$ must be 
larger than 3/4 \cite{massar,fuchs}. Since we only estimated 
the two lowest order contributions (to $p^2$ and $p^3$), the fidelity is 
also correct up to $p^2$ and $p^3$, and we write $F^{(2)}$ and $F^{(3)}$ 
respectively. Using Eqs.\ (\ref{fidelity}) and (\ref{ord2cas2}) we have:
\begin{equation}\label{2oinns}
 F^{(2)} = \frac{np_2}{p_1 [1+(5n-3)(1-\eta_c^2)]+np_2} \geq\frac{3}{4} \; ,
\end{equation}
\begin{equation}\label{2oeff}
 \Longrightarrow\qquad \eta_c^2 \geq \frac{(15n-6)p_1 - np_2}{(15n-9)p_1} \; .
\end{equation}
This means that in the limit of infinite detector cascading 
($n\rightarrow\infty$) and $p_1 = p_2$ the efficiency of the detectors must be 
better than 93.3\% to achieve non-post-selected quantum teleportation. When we 
have detectors with efficiencies of $98\%$, we need at least four detectors in 
the cascade to get unequivocal quantum teleportation. {\em The necessity of a 
lower bound on the efficiency of the detectors used in the cascade might seem
surprising}, but this can be explained as follows. Suppose the detector
efficiencies become smaller than a certain value $x$. Then upon a two-photon 
state entering the detector, finding only one click becomes more likely than 
finding a coincidence, and `wrong' events end up contributing to the output 
state. Eq.\ (\ref{2oeff}) places a severe limitation on the practical use of 
detector cascades in this situation.

In the experiment in Innsbruck, no detector cascade was employed and also the
$a_y$-mode was left undetected. The state entering Bob's detector therefore 
was (up to order $p^2$):
\begin{equation}\label{actual}
  \rho_{\text{out}} \propto \frac{p^2}{8} g_{uvc} \left[ 
  (3-\eta^2_c)|0\rangle\langle 0| + |\Psi_\theta\rangle
	\langle\Psi_\theta| \right] + O(p^3) \; .
\end{equation}
Remember that $p_1 = p_2$ since the experiment involves one source which
is pumped twice. The detector efficiency $\eta_c^2$ in the Innsbruck 
experiment was 10\% \cite{weincomm}, and the fidelity without detecting the 
outgoing mode therefore would have been $F^{(2)}\simeq 26 \% $ (conditioned 
only on successful Bell detection and state-preparation). This clearly 
exemplifies the need for Bob's detection. Braunstein and Kimble 
\cite{braunstein} predicted a theoretical maximum of 50\% for the 
teleportation fidelity, which was conditioned upon (perfect) detection of both 
the $a_x$- and the $a_y$-mode.

Rather than improving the detector efficiencies and using a detector cascade, 
Eq.\ (\ref{2oinns}) can be satisfied by adjusting the probabilities $p_1$ and 
$p_2$ of creating entangled photon-pairs \cite{braunstein}. From Eq.\ 
(\ref{2oinns}) we have
\begin{equation}
 p_1 \leq \frac{n}{3[1+(5n-3)(1-\eta_c^2)]} p_2 \; .
\end{equation}
Experimentally, $p_1$ can be diminished by employing a beam-splitter with a 
suitable reflection coefficient rather than a mirror to reverse the pump beam 
(see figure \ref{fig:innsbruck}). Bearing in mind that $\kappa$ is proportional
to the pump amplitude, the equation $p_i = 2\tanh^2(\kappa_i t)$ [see the 
discussion following Eq.\ (\ref{EPRdist2}) with $i=1,2$] gives a relation 
between the pump amplitude and the probability of creating a photon-pair. In 
particular when $p_2 = x p_1$:
\begin{equation}
  \frac{\tanh(\kappa_2 t)}{\tanh(\kappa_1 t)} = \sqrt{x} \; .
\end{equation}
Decreasing the production rate of one photon-pair source will increase the 
time needed to run the experiment. In particular, we have from Eq.\ 
(\ref{actual}) that
\begin{equation}
 p_2 \geq 3(3-\eta_c^2) p_1 \; .
\end{equation}
With $\eta^2_c=10\%$, we obtain $p_2 \geq 8.7 p_1$. Using Eq.\ (\ref{speed}) we
estimate that diminishing the probability $p_1$ by a factor 8.7 will increase 
the running time by that same factor (i.e., running the experiment about nine 
days, rather than twenty four hours).

The third-order contribution to the outgoing density matrix without cascading 
and without detecting the $a_y$-mode is
\end{multicols}

\noindent\rule{5cm}{.5pt}

\begin{eqnarray}\label{3oinns}
 \rho_{\text{out}} &\propto & 
  \frac{p_1}{8} g_{uvc} (4 - \eta_u^2 - \eta_v^2)\frac{1}{16} 
  \biggl[ 6p_1^2 (6 - 4\eta_c^2 + \eta_c^4)~|0\rangle\langle 0| \biggr. \cr
  && \qquad\biggl. + 2 p_1 p_2 (2 - \eta_c^2) \bigl( |\Psi_\theta\rangle
  \langle\Psi_\theta| + |\Psi^{\perp}_\theta\rangle\langle\Psi^{\perp}_\theta| 
  \bigr) + 8 p_1 p_2 (3 - \eta_c^2)\rho_1 + 12 p_2^2 \rho_2 \biggr] 
\end{eqnarray}
with
\begin{eqnarray}
 \rho_1 &=& \frac{1}{2} \left( |1,0\rangle\langle 1,0| + |0,1\rangle
	\langle 0,1| \right) \; ,\cr
 \rho_2 &=& \frac{1}{6} \biggl[ (2 + \cos 2\theta) |0,2\rangle\langle 0,2| +
	(2 - \cos 2\theta) |2,0\rangle\langle 2,0| + 2 |1,1\rangle\langle 1,1| 
	\biggr. \cr
	&& \qquad\left. + \frac{1}{2}\sqrt{2}\sin 2\theta \left( |2,0\rangle
	\langle 1,1| + 
	|1,1\rangle\langle 2,0| + |0,2\rangle\langle 1,1| + |1,1\rangle
	\langle 0,2| \right) \right] \; .
\end{eqnarray}

\hfill\rule{5cm}{.5pt}

\begin{multicols}{2}
We have explicitly extracted the state which is to be teleported
($|\Psi_\theta\rangle\langle\Psi_\theta|$) from the density matrix 
contribution 
$\rho_1$ (this is not {\em necessarily} the decomposition with the largest 
$|\Psi_\theta\rangle\langle\Psi_\theta|$ contribution). As expected, this term 
is less important in the third order than it is in the second. In the appendix 
it is shown that the $n$-photon contribution to the outgoing density matrix is 
always proportional to $p_2^n$.

The teleportation fidelity including the third-order contribution 
(\ref{3oinns}) can be derived along the same lines as (\ref{2oinns}). Assuming
that all detectors have the same efficiency $\eta^2$ and $p_1=p_2=p$, the 
teleportation fidelity up to third order is
\begin{equation}
 F^{(3)} = \frac{4+p(2 - \eta^2)^2}{4 (4 - \eta^2) + 
	   p (80 - 76 \eta^2 + 34 \eta^4 - 3 \eta^6)} \; .
\end{equation}
With $p=10^{-4}$ and a detector efficiency of $\eta^2 = 0.1$, this 
fidelity differs from (\ref{2oinns}) with only a few parts in ten thousand:
\begin{equation}\label{ratio}
 \frac{F^{(2)}-F^{(3)}}{F^{(2)}} ~\propto~ p ~\sim~ 10^{-4} \; .
\end{equation} 

On the other hand, let us compare two experiments in which the cascades have 
different detector efficiencies (but all the detectors in one cascade still 
have the same efficiency). The ratio between the teleportation fidelity with 
detector efficiencies $\eta^2_-$ and $\eta^2_+$ (with $\eta^2_-$ and $\eta^2_+$
the lower and higher detector efficiencies respectively) up to lowest order is
\begin{equation}
 \frac{F^{(2)}_{95\%}-F^{(2)}_{10\%}}{F^{(2)}_{95\%}} ~\propto~ \frac{\Delta
  \eta^2}{2-\eta^2_-} ~\sim~ 10^{-1} \; ,
\end{equation}
where $\Delta\eta^2$ is the difference between these efficiencies. This shows 
that detector efficiencies have a considerably larger influence on the 
teleportation fidelity than the higher-order pair production.

To summarise our results, we have found that detector cascading is only useful 
when the detectors in the cascade have near unit efficiency. In particular, 
there is a lower bound to the efficiency below which an increase in the number
of detectors in the cascade actually decreases the ability to distinguish 
between one or several photons entering the cascade. Finally, enhancement of 
the photon-pair source responsible for the entanglement channel relative to the
one responsible for the state preparation increases the time needed to run the
experiment by an order of magnitude. 

\section{Conclusions}\label{conclusions}

We studied the experimental realisation of quantum teleportation as performed 
in the Innsbruck experiment \cite{bouwmeester} including possible improvements 
suggested by Braunstein and Kimble to achieve a high non-post-selected fidelity
\cite{braunstein}. The creation of entangled photon-pairs using parametric 
down-conversion was analysed and we presented a discussion about the 
teleportation fidelity. Finally, we determined the usefulness of detector 
cascading and the slow-down of one down-converter relative to the other for 
the generalised experiment.

The difficulties of the Innsbruck experiment can be traced to the 
state-preparation (i.e., to the sources of the entangled photon-pairs, see
figure \ref{model}). In particular, there is a probability that the source 
responsible for creating entangled photon-pairs produces two pairs 
simultaneously. We studied these sources in some detail and have found that 
{\em photon-pairs created in a parametric down-converter are not independent 
of each other}. Employing two parametric down-converters therefore 
automatically probes the non-Poissonian structure of these sources.

The teleported {\em state} in the Innsbruck experiment is a mixture of the 
vacuum and a single-photon state. However, we cannot interpret this state as a 
low-efficiency teleported state, where sometimes a photon emerges from the 
apparatus and sometimes not. This reasoning is based on a particular partition 
of the outgoing density matrix, and this is not consistent with quantum 
mechanics (to our knowledge, this is the first instance where {\sc pef} leads 
to a different evaluation of an experiment). In section \ref{whatisqt} we 
showed how a high fidelity in the Innsbruck experiment could only be 
interpreted in a post-selected manner.

The interpretation of what quantum teleportation is, gives rise to different 
evaluations of the Innsbruck experiment. When one holds that the freely 
propagating output state of quantum teleportation should resemble the input 
state sufficiently closely (i.e., non-post-selected quantum teleportation), the
non-post-selected teleportation fidelity in the Innsbruck experiment should be 
at least $3/4$. This requirement was not met.  Nonetheless the Innsbruck 
experiment demonstrated {\em post-selected} quantum teleportation (i.e., 
teleportation conditioned on the detection of the outgoing state).

In the generalised version of the Innsbruck experiment ({\it \`a la} Braunstein
and Kimble) we have modelled a detector cascade in the state-preparation mode.
However, for the cascade to work, the detectors need to have near unit
efficiency. In particular, for infinite cascading the efficiency of the 
detectors should be at least ninety percent. Finite cascading requires even
higher detector efficiencies. This places a severe limitation on the practical 
use of detector cascades in this situation. Detector losses in the cascade 
have an immediate influence on the teleportation fidelity, yielding an effect 
which is much stronger than the higher order corrections due to multiple-pair 
creation (three pairs or more) of the down-converters. 

If the stability of the experimental setup can be maintained for a longer time
(the order of weeks), it is possible to slow down the down-converter 
responsible for creating the unknown input state. This can improve the 
fidelity up to arbitrary level. Nevertheless, we feel that our analysis 
demonstrates the definite benefits of single-photon detectors for such 
experiments or applications in the future. This technology currently requires 
roughly $6^{\circ}$K operating conditions.

\section*{Acknowledgements}

This research is funded in part by EPSRC grant GR/L91344.

\appendix

\section*{Cross-terms}

In this appendix we show that all the cross-terms of the density matrix in
Eq.\ (\ref{out}) must vanish. The density matrix consists of
several distinct parts: a vacuum contribution, a contribution due to
one photon in mode $d$, two photons, and so on. Suppose there are $n$
photon-pairs created in the whole system, and $m$ photon-pairs out of $n$ are 
produced by the second source (modes $c$ and $d$). The outgoing mode must then 
contain $m$ photons. Reversing this argument, when we find $m$ photons in the
outgoing mode the probability of creating this particular contribution must be
proportional to $p_1^{n-m} p_2^m$. Expanding the $n$-th order output state
into parts of definite photon number we can write
\begin{equation}
  \rho_{\text{out}}^{(n)} = \sum_{m=0}^{n-1} p_1^{n-m} p_2^m \rho_m^{(n)} \; ,
\end{equation}
where $\rho_m^{(n)}$ is the (unnormalised) $n$-th order contribution containing
all terms with $m$ photons. 

An immediate corollary of this argument is that all the cross-terms between
different photon number states in the density matrix must vanish. The 
cross-terms {\em are} present in Eq.\ (\ref{subout}), and we must therefore 
show that the partial trace in Eq.\ (\ref{out}) makes them vanish. Suppose 
there are $n$ photons in the total system. A cross-term in the density matrix
will have the form
\begin{equation}\nonumber
 | j, k, l, m \rangle_{auvd}\langle j', k', l', m' | \; ,
\end{equation}
with $m\neq m'$. We also know that $j+k+l+m=j'+k'+l'+m'=n$,
so that at least one of the other modes must have the cross-term property as 
well. Suppose $k$ is not equal to $k'$. Since we have 
Tr$[|k\rangle\langle k'|] = \delta_{k,k'}$, the cross-terms must vanish.

\end{multicols}


\begin{references}
\bibitem{note} The fidelity captures one particular feature of quantum
	teleportation very well and is extensively studied so far.
\end{references}
\end{document}